\documentclass[useAMS,usenatbib]{mn2e}

\title[Behaviour of dissipative accretion flows around black holes]
  {Behaviour of dissipative accretion flows around black holes}
\author[Santabrata Das]
  {Santabrata Das\thanks{sbdas@canopus.cnu.ac.kr}
  \newauthor 
  \\
  ARCSEC, 98 Gunja-Dong, Gwangjin-Gu,
         Seoul 143-747, Sejong University, South Korea.}
\date{\today}

\pagerange{\pageref{firstpage}--\pageref{lastpage}} \pubyear{2002}

\def\LaTeX{L\kern-.36em\raise.3ex\hbox{a}\kern-.15em
    T\kern-.1667em\lower.7ex\hbox{E}\kern-.125emX}

\usepackage{graphicx}
\input psfig.tex

\def\lsim{\mathrel{\hbox{\rlap{\hbox{\lower4pt\hbox{$\sim$}}}\hbox{$<$}}}}
\def\gsim{\mathrel{\hbox{\rlap{\hbox{\lower4pt\hbox{$\sim$}}}\hbox{$>$}}}}
\def \simeq{\lower.3ex\hbox{$\; \buildrel \sim \over - \;$}}

\begin{document}

\label{firstpage}

\maketitle

\begin{abstract}

We investigate the behaviour of dissipative accreting matter close to a black hole
as it provides the important observational features of galactic and extra-galactic
black holes candidates. We find the complete set of global solutions in 
presence of viscosity
and synchrotron cooling. We show that advective accretion flow can have 
standing shock wave and the dynamics of the shock is controlled by the dissipation 
parameters (both viscosity and cooling). We study the effective region of the 
parameter space for standing as well as oscillating shock. We find that shock 
front always moves towards the black hole as the dissipation parameters are 
increased.  However, viscosity and cooling have opposite effects in deciding 
the solution topologies. We obtain two critical cooling parameters that 
separate the nature of accretion solution.

\end{abstract}


\section{Introduction}

Accretion process around compact objects has been extensively studied during 
last three decades. Several attempts have been made for the development of the 
theory of accretion disc. The standard theory of thin accretion disc provides
a self-consistent solution of Keplerian disc model (\citealt{ss73}, hereafter SS73). 
Soon after the standard disc
model was proposed people realized that the disc are not Keplerian everywhere,
particularly in the radiation pressure dominated region (\citealt{le74},
\citealt{c96pr}). In addition, this model (SS73) was unsuccessful
to explain the origin of observed high energy radiation. A possible solution 
immediately came in the literature \citep{suti85}
that disc may possess a Compton cloud which must inverse Comptonize the Keplerian soft 
photons to produce the hard X-rays. Incidentally, \citet{cm95} and \citet{lmc98} 
showed by numerical simulation that angular momentum distribution purely depends
on the viscosity prescription under consideration and remains sub-Keplerian close 
to the black hole as flow must cross the horizon super-sonically \citep{c90a}. 
Recent observations do support the fact that sub-Keplerian matter may be present 
in accretion disc (\citealt{shms01}, \citealt{shs02}).

In an earlier study, Chakrabarti (\citealt{c89}, \citealt{c90a})
extensively discussed the transonic properties of inviscid, polytropic 
flows and showed that standing shock waves can form for a large region of parameter 
space spanned by the specific energy and specific angular momentum of the flow.
Various authors (\citealt{nh94}, \citealt{yk95}, \citealt{ly97}, \citealt{gl04}, 
\citealt{ft04})
found the existence of standing shock waves in different astrophysical contexts.
In an accretion process around a black hole, flow is sub-sonic at a large distance 
and crosses the horizon super-sonically. Thus, black
hole accretion is necessarily transonic. For a given set of input parameters, namely
specific energy and specific angular momentum, the flow passes through the outer sonic 
point $(x_{out})$ and remains super-sonic as it proceeds inwards. At a distance 
$x_s~(< x_{out})$,  flow is virtually stopped due to the resistance offered by the 
centrifugal barrier and consequently a 
shock is formed. The post-shock flow momentarily becomes sub-sonic and 
flow temperature increases in this region 
as the kinetic energy of the flow is converted in to thermal energy. Subsequently, flow
picks up its velocity as it approaches black hole horizon and eventually
crosses the horizon supersonically after passing through the inner sonic point $(x_{in})$.

The first global and fully self-consistent solution of viscous transonic flows including
advection in the optically thin and thick limit was  obtained by Chakrabarti 
(\citealt{c90a} and references there in) 
where it was assumed that Keplerian flow at the outer edge 
can become sub-Keplerian at the inner
part of the disc. Later, \citet{cd04} 
presented a complete classification
of global solutions of viscous transonic flow according to solution topologies and
identified the region in the parameter space where the flow possesses multiple sonic points.
Within this region, if the viscosity parameter is below its critical value 
$(\alpha_{\Pi} < \alpha_{\Pi s})$,
there are solutions where Rankine-Hugoniot shock conditions (\citealt{ll59}, hereafter RHCs) are  
satisfied and standing shock is formed due to centrifugal barrier (\citealt{c90a}, 
\citealt{c96b}, \citealt{cd04}). When viscosity is increased beyond a critical
value, shocks disappear (see Fig. 12 of \citealt{cd04}).  
Of course, the viscosity parameter strongly 
depends on the model parameters such as sonic point location and angular momentum at the
inner boundary. The importance of critical viscosity is being reconsidered by \citet{gl04} 
where a `qualitative' estimate of critical viscosity for standing shock 
($\alpha_{\Pi s} \leq 0.5$) was reported. Furthermore, many authors pointed out that shocks
may undergo either radial or vertical or non-axisymmetrical oscillations 
(\citealt{mtk99}, \citealt{gf03}) although they are generally never ruined by these oscillations.
In \citet{dc04}, 
Bremsstrahlung cooling also added keeping in mind that it is a very inefficient 
cooling process \citep{cc00}
and complete set of global solutions 
of viscous transonic flow with and without shocks were presented.  The hot and
dense post shock flow which basically acts as a `boundary layer' of the black
holes could be the natural site of the hot radiation in the 
accretion disc and is considered to be a powerful tool
in understanding the important astrophysical phenomena such as the spectral
properties of black holes (\citealt{ct95}, \citealt{ctke96}, \citealt{etc96},
\citealt{mc05}),
the source of quasi-periodic oscillation (QPO) of the hard X-rays from the black hole 
candidates (\citealt{msc96}, \citealt{rcm97}, \citealt{cam04}) and 
the generation of accretion-powered relativistic bipolar outflows/jets 
(\citealt{dcnc01} and references there in).
Recently, \citet{cm06} 
reported that
this boundary layer could be responsible for the spectral state
transition in at least some of the black hole candidates, such as Cyg X1.
Also, \citet{fetal00} and \citet{dmr00} 
showed that outflow could be produced from the same region that emits the 
Comptonized photons. A globally complete Inflow-Outflow solution of viscous accretion
flow is also found by \citet{cd06}. 

Being motivated by the above arguments, in the present paper, we wish to 
follow up the stationary, axisymmetric, viscous accretion solutions around 
a Schwarzschild black hole in presence of Synchrotron cooling. Such 
dissipative viscous accreting flow have not yet been explored in the
literature. The space-time 
geometry around a non-rotating black hole is satisfactorily described by 
pseudo-Newtonian potential \citep{pw80}. 
We consider similar viscosity prescription as in \citet{c96b}. 
We calculate all relevant
dynamical and thermodynamical flow variables and extensively study the
dependence of such variables on the the flow parameters like
specific energy, specific angular momentum and the dissipation 
parameters (Synchrotron cooling and/or viscosity parameters). 
We identify the solution topologies which are essential for shock formation.
In the present study, we concentrate on the non-dissipative shocks
that preserve the energy across it. In a generalized accretion flow, 
cooling reduces the energy of the flow while viscosity not only tends 
to heat the flow but lowers the flow angular momentum as flow accretes. 
Thus the effect of viscosity and Synchrotron cooling on the dynamical
structure of the accretion flow are expected to be different.
Since shock has observational 
consequences as mentioned above, it is, therefore, pertinent to know how 
the dynamics of shock is controlled by the dissipation parameters. In 
particular, whether shock front moves inward or outward if the dissipation
is increased. In the present paper, we attempt to discuss this issue extensively.
Moreover, we find that shock waves, standing or oscillating, are still formed 
even when at very high dissipation limit. In this paper, we provide a useful
formalism to study the dissipative transonic accretion flow which is valid 
for a wide range in accretion rates. 

We arrange the paper in the following way: in the next section, we present
the governing equations. In \S 3, we perform
sonic point analysis. In \S 4, we study the global solution topologies 
and classify the parameter space for solutions having multiple sonic 
points as a function of cooling parameter. In \S 5, we study the dynamics
of shock and dependence of shock properties on the the flow parameters.
In \S 6, we classify the the parameter space as a function of cooling
parameter in terms of whether shocks, standing or oscillating, will 
form or not. In \S 7, we quantify the critical cooling parameters. Finally,
in \S 8, we make concluding remarks.

\section{Basic hydrodynamic equations}

We begin with a stationary, thin, viscous, axisymmetric accretion 
flow on to a Schwarzschild black hole. In this paper, we consider 
pseudo-Newtonian potential introduced by \citet{pw80} 
to solve the problem instead of using full general relativity, 
which allows us to use the Newtonian
concept at the same time retaining all the salient features of the
space-time geometry around a non-rotating black hole. The governing 
equations for the accreting flow are written on the equatorial plane 
of the accretion disc.
The flow 
equations are made dimensionless by considering unit of length, time and the 
mass as $r_g=2GM_{BH}/c^2$,  $2GM_{BH}/c^3$ and $M_{BH}$ respectively where, $G$
is the  gravitational constant, $M_{BH}$ is the mass of the black hole and 
$c$ is the velocity of light respectively. Henceforth, all the flow variables are expressed
in geometrical units.

In the steady state, the hydrodynamics of the axisymmetric accreting matter around a 
non-rotating black hole in the pseudo-Newtonian limit is given by \citep{c96b},

\noindent (i) the radial Momentum equation :
$$
\vartheta \frac {d\vartheta}{dx}+\frac {1}{\rho}\frac {dP}{dx}
-\frac {\lambda^2}{x^3} + \phi (x)=0,  
\eqno{(1)}
$$

\noindent (ii) the mass flux conservation equation :
$$
\dot M = \Sigma \vartheta x
\eqno{(2)}
$$

\noindent (iii) the angular momentum conservation equation :
$$
\vartheta \frac {d\lambda(x)}{dx}+\frac{1}{\Sigma x}
\frac {d}{dx}\left( x^2 W_{x\phi}\right)=0,
\eqno{(3)}
$$

and finally,

\noindent (iv) the entropy generation equation :
$$
\Sigma \vartheta T \frac {ds}{dx}=\frac{h\vartheta}{\gamma-1}
\left(\frac{dP}{dx} -\frac{\gamma P}{\rho}\frac{d\rho}{dx}\right) = Q^+-Q^- .
\eqno{(4)}
$$

The local variables $x$, $\vartheta$, $\rho$, $P$ and $\lambda$ 
in the above equations are the radial distance, radial velocity, density,
isotropic pressure and the angular momentum of the flow respectively. 
Here, $\phi(x) = -\frac{1}{2}(x-1)^{-1}$ denotes the pseudo-Newtonian 
potential and $\gamma$ is the adiabatic index of the flow. Furthermore, 
$s$ and $T$ are the specific entropy and the local temperature of the flow, 
$\Sigma$ is the vertically integrated density and $W_{x\phi}$ is the 
viscous stress. Here, $Q^+$ and $Q^-$ represent the energy gained and 
lost by the flow respectively, and ${\dot M}$ is the mass accretion rate. 
In our model, the accreting flow is assumed to be in hydrostatic 
equilibrium in the vertical direction. 
The half thickness of the disc is then obtain by equating the vertical 
component of the gravitational force and the pressure gradient force as,

$$
h(x)=ax^{1/2}(x-1).
\eqno{(5)}
$$
where, $a$ denotes the adiabatic sound speed defined as 
$a=\sqrt {\gamma P/\rho}$. In this paper, we use the similar
viscosity prescription of \citet{c96b}. 
In this prescription, viscous stress remain continuous across 
the axisymmetric shock wave
for flows with significant radial motion.

Now we simplify the entropy equation (Eq. 4) as:

$$
\frac{\vartheta}{\gamma-1}\left[ \frac{1}{\rho} \frac{dP}{dx}
-\frac{\gamma P}{\rho^2}\frac{d\rho}{dx}\right]=\frac{Q^--Q^+}
{\rho h}=\Lambda-\Gamma,
\eqno(4a)
$$
where, $\Gamma (=Q^{+}/\rho h)$ denotes the energy gained due to the viscous 
heating and is obtained as,

$$
\Gamma=\frac{-\alpha_{\Pi} I_n x}{\gamma}\left(ga^2+
\gamma\vartheta^2\right)\frac{d\Omega}{dx}.
\eqno(6)
$$  

Here, $\alpha_{\Pi}$ is the viscosity parameter and $\Omega(x)$ is the
angular velocity of the accreting matter at a radial distance $x$. The
polytropic index is denoted by $n =(1-\gamma)^{-1}$. $I_n$ and $I_{n+1}$
come from the vertically averaged density and pressure \citep{mk84} and
$g=I_{n+1}/I_n$. 
The term $\Lambda (=Q^{-}/\rho h)$ represents the energy loss by the flow.
In the present analysis, we ignore energy loss due to Bremsstrahlung cooling
as it is not very efficient cooling process \citep{cc00} and include 
synchrotron cooling only. Indeed, magnetic field is ubiquitous inside the 
accretion disc and therefore, the ionized flow in turn should emit 
synchrotron radiation that may cool down the accreting flow significantly.
Since the satisfactory description of magnetic field inside the accretion disc is not 
well understood, we thus assume only random or stochastic 
magnetic field. Such magnetic field may or may not be in equipartition 
with the accretion flow. We define a parameter $\eta$ as the ratio of 
the magnetic pressure and the thermal pressure of the flow under 
consideration and is given by,

$$
\eta=\frac{B^2 \mu m_p}{8\pi \rho k_B T_p}
\eqno(7)
$$
where, $B$ denotes the magnetic field strength, $k_B$ is the Boltzmann 
constant, $\mu$ is the mean molecular 
weight and $m_p$ is the mass of the proton respectively. In general,
$\eta \lsim 1$ which ensures that magnetic fields definitely remain
confined within the disc \citep{mc05}.
While obtaining Eq. (7), we consider the gas is
assumed to obey an ideal equation of state.  
In an electron-proton plasma, we use equipartition 
magnetic field (Eq. 7) to account for the non-dimensional synchrotron 
cooling effect which is obtained in a convenient form as \citep{shte83}:

$$
\Lambda=\frac{S a^5}{\vartheta x^{3/2}(x-1)},
\eqno(8)
$$
with

$$
S=7.5424 \times \frac{\eta \beta {\dot M} \mu^2 e^4}{m_e^3\gamma^{5/2}}
\frac{1}{2GM_{BH}c^3},
\eqno(9)
$$
where, $e$ and $m_e$ represent the charge and mass of 
the electron respectively. We estimate electron temperature
from the expression $T_e=\sqrt{m_e/m_p}T_p$ \citep{cc02} and use it
to obtain Eq. (8). Here, $\beta$ is a dimensionless {\it cooling parameter} 
that controls the efficiency of cooling. When $\beta \rightarrow 0$, the flow
is heating dominated as it cools most inefficiently. 
In the present paper, we consider $\gamma = 4/3$, $\eta = 0.1$ and 
$\dot {M}= 0.233 \dot {M}_{Edd}$ for $10M_{\sun}$ black hole, until otherwise stated. 

\section{Sonic point analysis}

We solve Eqs.(1-3, 4a) following the standard method of sonic point analysis
\citep{c89}. We calculate the radial velocity gradient as:
$$
\frac {d\vartheta}{dx}=\frac{N}{D},
\eqno(10)
$$
where, the Numerator $N$ is given by,

$$
N =\frac {Sa^5}{x^{3/2}(x-1)}
+\frac {\alpha^2_\Pi I_n  (a^2g+\gamma \vartheta^2)^2}{\gamma^2 x}
+\frac {\alpha^2_\Pi g I_n a^2(5x-3)(a^2g+\gamma\vartheta^2)}{\gamma^2 x(x-1)}
$$
$$
-\left[ \frac {\lambda^2}{x^3}-\frac {1}{2(x-1)^2}\right]
\left[\frac {(\gamma+1)\vartheta^2}{(\gamma-1)} - 
\frac{2\alpha^2_\Pi g I_n (a^2g+\gamma\vartheta^2)}{\gamma} \right]
$$
$$
-\frac {\vartheta^2 a^2(5x-3)}{x(\gamma-1)(x-1)}
-\frac {2\lambda \alpha_\Pi I_n \vartheta (a^2g+\gamma\vartheta^2)}{\gamma x^2}
\eqno{(10a)}
$$

and the denominator $D$ is,
$$
D = \frac {2a^2\vartheta}{(\gamma-1)}-\frac {(\gamma+1)\vartheta^3}
{(\gamma-1)}
+\frac{\alpha^2_\Pi I_n \vartheta(a^2g+\gamma\vartheta^2)}{\gamma}
\left[ (2g-1)-\frac {a^2g}{\gamma\vartheta^2}\right] .
\eqno{(10b)}
$$

The gradient of sound speed is obtained as:

$$
\frac{da}{dx}=\left( \frac{a}{\vartheta} - \frac{\gamma \vartheta}{a} \right)
\frac{d\vartheta}{dx} + \frac{\gamma}{a}\left[ \frac {\lambda^2}{x^3}-\frac {1}{2(x-1)^2}\right]
+\frac{(5x-3)a}{2x(x-1)}
\eqno(11)
$$
and similarly, the gradient of angular momentum is calculated as:

$$
\frac{d\lambda}{dx}=\frac{\alpha_{\Pi}(a^2g+\gamma\vartheta^2)}{\gamma \vartheta}
-\frac{\alpha_{\Pi} x (a^2g- \gamma\vartheta^2)}{\gamma \vartheta^2}\frac{d\vartheta}{dx}
+\frac{2 \alpha_{\Pi} axg }{\gamma \vartheta}\frac{da}{dx}
\eqno(12)
$$

At the outer edge of the accretion disc, matter starts to accrete with 
almost negligible radial velocity and enters into the black hole 
with velocity of light. This indicates that `radial velocity 
gradient' should be always real and finite to maintain the accretion 
flow smooth along the streamline. However, Eq. (10b) indicates that 
there may be some points where denominator ($D$) vanishes. Since the 
flow is smooth everywhere along the streamline, the point where 
denominator tends to zero, the numerator ($N$) must also vanish 
there. The point where both the numerator and denominator vanish 
simultaneously is called as critical point or sonic point. Setting 
$D=0$, one can easily obtain the expression for the Mach number 
$(M=\vartheta/a)$ at the sonic point as,

$$
M(x_c) =\sqrt {\frac{-m_2 - \sqrt{m^2_2-4m_1 m_3}}{2m_1}},
\eqno(13)
$$
where,
$$ 
m_1=\gamma^2\left[\alpha^2_{\Pi} I_n (2g-1)(\gamma-1) - (\gamma+1)\right]
$$
$$
m_2=2\gamma\left[\gamma+\alpha^2_{\Pi} I_n g (g-1)(\gamma-1) \right]
$$ 
$$
m_3=-\alpha^2_{\Pi} I_n g^2 (\gamma-1)
$$ 

In the weak viscosity limit, the Mach number at the sonic point becomes,

$$
M(x_c) = \sqrt{\frac {2}{\gamma+1}}
\eqno(13a)
$$  
This result is exactly same as \citet{c89}. This indicates that Mach number at the
sonic point in presence of different cooling process remains same as in the
case for the non-dissipative accretion flow.
We obtain an algebraic equation for sound speed by using another sonic point
condition $N=0$ which is given by, 

$$
F({\mathcal E}_c, \lambda_c, x_c)={\mathcal A}a^3(x) + {\mathcal B}a^2(x)
+ {\mathcal C}a(x) +{\mathcal D}= 0 ,
\eqno(14)
$$
where,
$$
{\mathcal A}=\frac{S}{x^{3/2}(x-1)},
\eqno(14a)
$$
$$
{\mathcal B} =  \frac {\alpha^2_\Pi I_n (g+\gamma M^2)^2}
{\gamma^2 x}+\frac {\alpha^2_\Pi I_n g (5x-3)(g+\gamma M^2)}
{\gamma^2 x(x-1)}
$$
$$
-\frac{M^2(5x-3)}{x(\gamma-1)(x-1)} ,
\eqno(14b)
$$

$$
{\mathcal C} = -\frac {2\lambda \alpha_\Pi I_n M (g+\gamma M^2)}{\gamma x^2},
\eqno(14c)
$$
and

$$
{\mathcal D} = -\left[ \frac {\lambda^2}{x^3}-\frac {1}{2(x-1)^2}\right]
\left[\frac {(\gamma+1) M^2}{(\gamma-1)}-\frac{2\alpha^2_\Pi g I_n (g+\gamma M^2)}{\gamma} \right].
\eqno(14d)
$$

We solve Eq. (14) analytically \citep{as70} and obtain the sound speed
at the sonic point by knowing the flow parameters. Depending on the 
input parameters, a non-dissipative
flow may have maximum of four sonic points \citep{dcc01}. Among them, one
always lies inside the horizon and the others, if they at all exist,
remain outside the horizon. In our present analysis, we continue to use similar
concept to obtain dissipative accretion solutions \citep{cd04}. 
The nature of sonic points and its detailed properties could be 
easily understood with the extensive study of Eq. (10) when the flow 
parameters are known. Infact, nature of sonic points depends on the value
of velocity gradients at the sonic point. In general,
$d\vartheta/dx$ possesses two values at the sonic point.
One of them is for accretion flow and the other is valid for wind. 
When both the derivatives are real and is of opposite signs, the sonic 
point is saddle type or X-type. Nodal type sonic point is obtained when
both the derivatives are real and is of same sign. If the derivatives
are complex, the sonic point belongs to the class of spiral type or O-type.
In the
astrophysical context, saddle type sonic point has some special importance 
since transonic flows usually pass through it. In addition, in order to form
a standing shock, the flow must have more than one saddle type sonic points 
among them the closest and furthest values corresponds to inner and outer
X-type sonic points. The O-type sonic point which lies between the two 
saddle type sonic point is unphysical in the sense that no stationary 
solution passes through it.

\section{Solution Topologies}

To obtain a complete solution, one needs to supply the boundary values 
of energy, accretion rate and angular momentum of the flow
for a given viscosity ($\alpha_{\Pi}$) and cooling parameter ($\beta$)
and integrates Eqs.(10, 11, 12) simultaneously with the help of 
Eqs. (10a, 10b). Since
black hole solution is necessarily transonic, flow must pass through the sonic points.
In addition, flow presumably originates from the Keplerian disc ($x_{Kep}$) and it
must carry angular momentum $\lambda(x_{Kep})$, a part of which is transported due
to viscosity outwards and the other part is advected into the black hole
through the sonic point. The distribution of the angular momentum 
from the inner edge of the disc to the $x_{Kep}$ is obtained from Eq. (12).
Incidentally, the acceptable range of angular momentum at the inner edge of 
the disc and the inner sonic point location of the accreting solution is very small 
(say, $1.5 \lsim \lambda_{in} \lsim 2$ and $2 \lsim x_{in} \lsim 4$, 
\citealt{c89}, \citealt{cd04}). We, thus, choose inner sonic point location $x_{in}$ and 
the angular momentum at the inner sonic point $\lambda(x_{in})$ 
(hereafter, we denote it as $\lambda_{in}$) 
as input parameters instead of the aforementioned energy and the angular 
momentum flow at the outer edge of the disc. We once integrate Eqs.(10, 11, 12)
inwards upto the horizon and again outwards to a large distance 
$(\sim x_{Kep})$ to
obtain a complete transonic solution.  A comprehensive
study of viscous transonic flow is already presented by \citet{cd04} and
therefore, in the current paper, we mainly concentrate on the properties of 
accreting solution in presence of synchrotron cooling, until otherwise stated.

\subsection{Behavior of Global Solutions}

In order to obtain a shock solution around a black hole, accreting flow must 
possess multiple saddle type sonic points. Shock joins different solution 
branches, one passes through the inner sonic point and the other passes 
through the outer sonic point. Thus, it is important to understand
the nature of solution trajectories in presence of energy dissipation. 

\begin{figure}
\begin{center}
\vskip 0.5cm
\includegraphics[height=3cm,width=8cm]{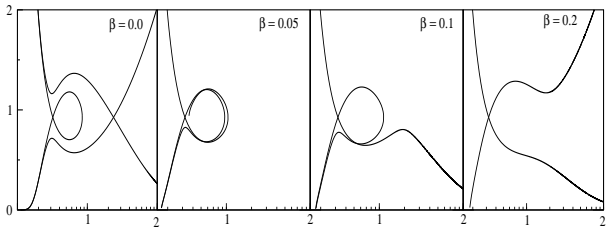}
\caption{Global solution topologies in presence of Synchrotron cooling.
In each boxes, variation of Mach number is plotted with logarithmic radial
distance. Flow parameters are $x_{in}=3.107$, $\lambda_{in} = 1.6$. Cooling
parameters are marked in the figure.
}
\end{center}
\end{figure}

Fig. 1 shows examples of global solutions of inviscid accretion flow passing through 
the inner sonic point. Each box depicts Mach number ($M=\vartheta/a$) variation 
as a function of the logarithmic radial distance.
We write the cooling parameter $\beta$ in each box
(marked as 0.0, 0.05, 0.1 and 0.2). 
The other flow parameters for Fig. 1 are chosen as $x_c=3.107$, $\lambda_{in}= 1.6$
and $n = 3$ respectively. 
Here, all the sonic points are saddle type, and the solutions may or may not have
shock depending on its global solution topologies. In the first column, the
flow is dissipation free and the sonic points are uniquely determined as
energy and angular momentum are conserved all throughout. Accreting solutions 
of this kind deviate from the Keplerian disc at the outer part of the disc
and enter into black hole super-sonically  with the radial velocity equal to the
velocity of light after crossing the outer sonic point.
Another possibility is that super-sonic flow may suffer shock transition at 
sub-sonic branch, if shock conditions hold (see \S 5) and crosses the black 
hole horizon after passing through the inner sonic point as it has to satisfy
the `super-sonic' inner boundary condition at the horizon.
In the second column, cooling is incorporated and the flow topologies passing 
through the inner sonic point start to open up. The outer sonic point, if it 
exits, can not be determined unless a shock is formed as energy of the flow 
is not conserved here. When 
cooling is increased further, the flow topologies are completely opened up (third and
fourth column) and they leave the accretion shock regime. In this particular situation,
flow deviates from the Keplerian disc far away from the black hole and enters into it
straight away through the inner sonic point. So, for a given set of flow parameters,
there must be a critical cooling parameter $(\beta_{cc})$ for which closed topologies
passing through the same inner sonic point become open topologies. An exhaustive 
discussion of critical cooling parameter will be presented in \S 7.

\subsection{Description of the parameter space}

\begin{figure}
\begin{center}
\vskip -1.2cm
\includegraphics[height=10cm,width=10cm]{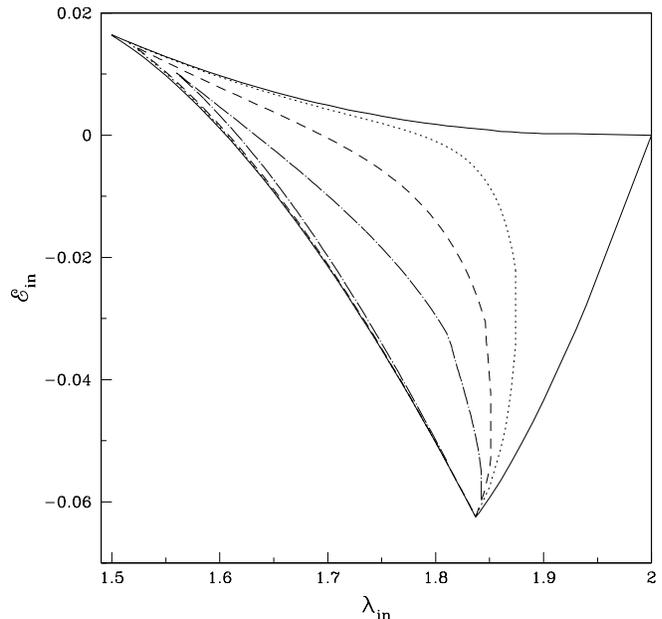}
\vskip -0.7cm
\caption{\small Classification of parameter space for closed topologies passing through the 
inner sonic point. The region bounded by different curve is obtained by using
different cooling factor [$\beta = 0.0 $ (solid), $\beta = 0.00787 $ (dotted), 
$\beta = 0.0787 $ (dashed) and $\beta = 0.393 $ (dot-dashed)]. Notice that, parameter 
space gradually shrinks with the increase of cooling factor. 
}
\end{center}
\end{figure}

In order to understand the dissipative accretion flow, here, we classify the parameter
space as a function of the cooling parameter in the $\mathcal {E}_{in}$-$\lambda_{in}$ 
plane where $\mathcal {E}_{in}$ denotes the energy of the flow at the inner
sonic point $(x_{in})$. Infact, in a cooling dominated inviscid accretion flow, 
angular momentum remain conserved along the streamline of the flow. In Fig. 2, we 
have identified the region of the parameter space for closed spiraling
accretion solution which passes through the inner sonic point as in panel 1-2 of Fig. 1.
Accretion flows of this kind possess multiple sonic points and flow may suffer shock
transition if the outer sonic point exits. The shock would be stationary
or oscillating depending on whether the Rankine-Hugoniot relation is satisfied or not.  
The region bounded by the solid curve is obtained for dissipation free ($\beta = 0$)
accretion flow and it coincides with the result of \citet{c89}. The dotted, dashed and the 
dot-dashed regions are obtained for cooling parameters $\beta = 0.00787, 0.0787$ 
and $0.393$ respectively.
For higher $\beta$, the parameter space for closed topologies passing through the inner sonic
point reduces in both lower and higher angular momentum sides. This is not surprising, since
an increase of cooling induces damping inside the flow which reduces the number the sonic
points for the same set of initial parameter as seen in Fig. 1. With the rise of $\beta$,
the effective region of parameter space for the closed topologies gradually shrinks and 
disappears completely when the critical cooling $(\beta_{cc})$ is reached.  

\section{Shock Solutions and classification of parameter space}

The existence of shock waves in accretion flow has been reported in many astrophysical
context and the thermodynamical properties of shock waves like its location, strength and
compression ratio could be determined exactly by using the RHCs. 
In the present study, we continue to use similar viscosity prescription as in 
\citet{c96b} that maintains viscous
stress $(W_{x \phi})$ to be continuous across the shock in presence of significant radial
motion.
We compute the shock locations following an useful technique first presented by 
\citet{cd04} for dissipative accretion flows.

\subsection{Shock Dynamics and Shock properties}

\begin{figure}
\begin{center}
\vskip -1.2cm
\includegraphics[height=10cm,width=10cm]{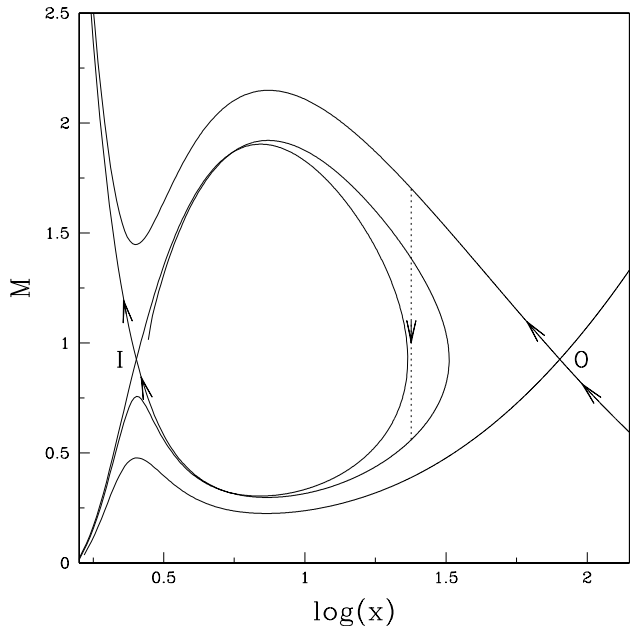}
\vskip -0.7cm
\caption{\small A complete accretion solution topology drawn with standing shocks ($x_s = 23.83$).
See text for details.
}
\end{center}
\end{figure}

In Figure 3, we present a complete solution of vertically averaged flow 
where a shock wave connects two solution
branches---one passing through the outer sonic point (O) and the other passing through the
inner sonic point (I). In this figure, we plot the Mach number variation with the logarithmic
radial distance. The flow parameters are chosen as  $x_{in}=2.5308, 
\lambda_{in}=1.725$ and $\beta = 0.0157$ respectively. Sub-sonic accretion flow 
crosses the outer sonic point at $x_{out}=80.07$ and makes discontinuous transition 
($shock$) at $x_s= 23.82$ before entering into the black hole.
The shock location is shown by vertical dotted line.
The exact location of the shock is obtained by solving Rankine-Hugotinot relations.
In the present study, shock is assumed to be thin and non-dissipative.
In addition, we ignore the presence of any excess sources of torque at the shock, that 
maintains the continuity of angular momentum across it. At the shock, entropy is 
generated which is subsequently advected into the black hole.

\begin{figure}
\begin{center}
\vskip -1.2cm
\includegraphics[height=10cm,width=10cm]{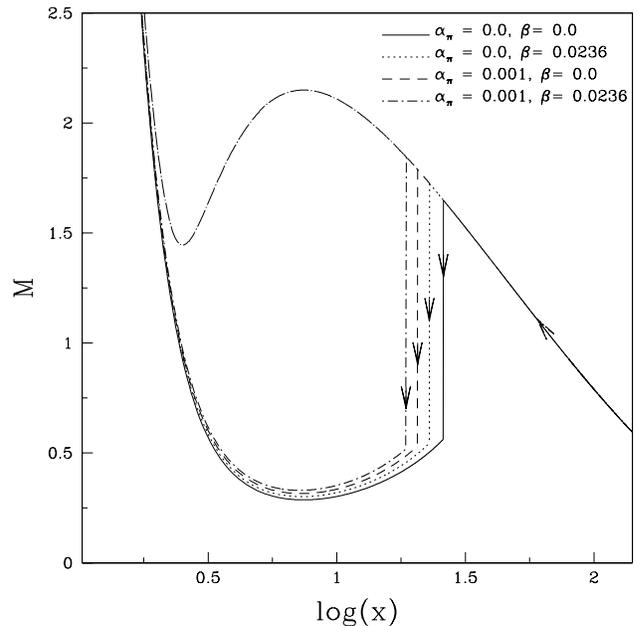}
\vskip -0.7cm
\caption{\small Plot of Mach number with the logarithmic radial distance.
Flows with same energy $\mathcal {E}_{inj} = 3.3663 \times 10^{-3}$ and 
angular momentum $\lambda_{inj} = 1.725$ at the outer edge $x_{inj}=145$
are injected with different dissipation parameters. Dissipation parameters
are marked in the figure. The corresponding shock locations are indicated 
by the vertical lines [$x_s = 25.99$ (solid), $x_s = 20.6$ (dashed), 
$x_s = 22.94$ (dotted) and $x_s = 18.62$ (dot-dashed),]. 
}
\end{center}
\end{figure}

In Fig. 4, we have shown how the shock location (indicated by the vertical lines)
changes with the dissipation parameters ($\alpha_{\Pi}$ and/or $\beta$) for a given
set of initial flow parameters. We inject matter sub-sonically 
at the outer edge of the disc $x_{inj}=145$ 
with local energy $\mathcal {E}(x_{inj}) = 3.3663 \times 10^{-3}$
and angular momentum $\lambda(x_{inj}) = 1.725$. 
(hereafter, we denote them as $\mathcal {E}_{inj}$ and $\lambda_{inj}$ respectively). 
The sub-sonic flow 
first passes through the outer sonic point and becomes super-sonic. The 
Rankine-Hugoniot condition are satisfied here and eventually a shock is formed. The solid 
vertical line represents the shock location ($x_s = 25.99$) for non-dissipation
($\alpha_{\Pi} =0$ and $\beta = 0$) flow. As viscosity is turn on, shock front 
($x_s = 20.6$) moves forward  denoted by the 
dashed vertical line. 
Due to viscosity, accreting flow losses angular momentum causing the 
reduction of centrifugal pressure and at the same time energy of the flow
increases. At the shock, total pressure
must be continuous and thus, the dynamics of the shock are mainly
controlled by the resultant pressure across it. 
Since the shock moves in and as $\lambda(x)$ get reduced 
along the viscous flow, therefore, we can conclude that centrifugal force 
is the primary cause for shock formation.
When synchrotron cooling is effective in an inviscid accretion flow, shock
location ($x_s = 22.94$) again proceeds towards the horizon depicted by 
vertical dotted line. In the pre-shock region, density and temperature are 
low and therefore, the effect of cooling is negligible (see below). The situation is 
completely opposite in the post-shock region and cooling reduces the post-shock 
pressure causing the shock to move forward further. This indicates that thermal
pressure is also not less important in determining the final shock location.
In a generalized accretion flow where both viscosity and cooling are present, shock
location is predicted at $x_s = 18.62$ for the same set of input parameter
as above and is denoted by vertical dot-dashed line. 
In this particular case, shock front is shifted significantly
due to combined effects of viscosity and cooling. When the mass loss from the viscous disc
\citep{cd06} is considered, shock is seen to form even closer to the black hole as some
part of the accreted matter is ejected as outflows which reduces the post-shock
pressure drastically.

\begin{figure}
\begin{center}
\vskip -4.0cm
\includegraphics[height=10cm,width=10cm]{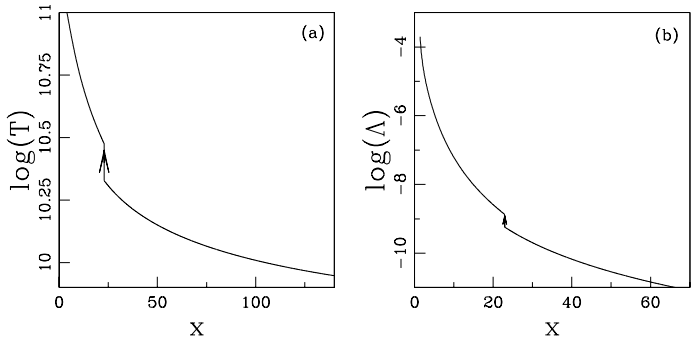}
\vskip -1.5cm
\caption{\small The variation of (a) Flow temperature and (b) total
Synchrotron loss (in geometric units) with radial distance.
}
\end{center}
\end{figure}

In Fig. 5a, we have plotted the logarithmic proton temperature
$(T=\mu m_p a^2/\gamma k_B)$  of a weakly viscous flow 
as a function of the radial
distance for the same set of input parameters as in Fig. 4 with $\beta = 0.0236$. 
Clearly Fig. 5a indicates that post-shock temperature is much higher compared to 
the pre-shock temperature. This is because shock compresses the flow and makes it
hotter. In particular, at the shock, flow temperature rises sharply
as the kinetic energy of the flow converted to the thermal energy there. In Fig. 5b,
we show how the total synchrotron loss (in logarithmic scale with geometric unit) 
vary with the radial 
distance for the same set of flow parameters as in Fig 5a. The net energy loss is
negligible in the pre-shock region while it increases significantly as the post-shock flow
advected towards the black hole. 

\begin{figure}
\begin{center}
\includegraphics[height=10cm,width=10cm]{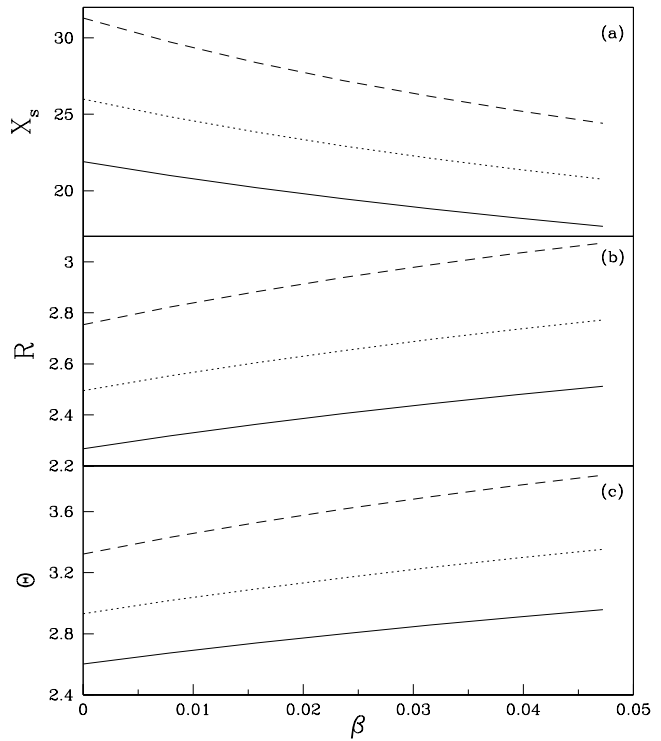}
\caption{\small (a) Variation of shock location with the cooling factor when the
flows with identical angular momentum and energy are injected from a fixed outer boundary.
Solid, dotted and dashed curves are drawn for flow with angular momentum 
$\lambda_{in} = 1.7, 1.725$ and $1.75$. Shock forms further for higher angular momentum
flow suggests that shock are mainly centrifugally driven. In all cases, shock moves 
closer to the black hole as the cooling factor is increased.
(b) Variation of compression ratio $(R=\Sigma_+/\Sigma_-)$ 
with the cooling factor for same set of
parameters as (a). Subscripts ``+'' and ``-'' refer to quantities before
and after shock. Compression ratio monotonically increases
in all the cases as the cooling factor is increased.
(c) Variation of shock strength $(\Theta=M_-/M_+)$ with the 
cooling factor for same set of
parameters as (a). Shock strength gradually increases
in all the cases as the cooling factor is increased.
}
\end{center}
\end{figure}

Fig. 6a compares the shock locations as a function of the cooling parameter 
$\beta$ for different values of angular momentum $\lambda_{inj}$. In all the 
cases, the matter is injected at the
outer edge of the disc $x_{inj}=145$. The solid curve is for $\lambda_{inj}= 1.7$ and 
the dotted and dashed curve are for $\lambda_{inj}=1.725$ and $\lambda_{inj}=1.75$
respectively. The 
corresponding local energies at the injection point are $\mathcal {E}_{inj}= 
2.5156 \times 10^{-3}$ (solid), $3.3663 \times 10^{-3}$ (dotted) and 
$4.3474 \times 10^{-3}$ (dashed) respectively. This example shows that
stable shocks can form for a wide range of cooling parameter. For a 
given local energy and angular momentum of the flow at the injection 
point, shock location decreases with the increase of the cooling parameter 
as flow losses energy as it accretes. When cooling is above its critical 
value ($\beta > \beta_{cs}$), 
shock ceases to exist as the RHCs are not 
satisfied there. An important point to note is that the shock forms at larger 
radii for higher angular momentum as the centrifugal barrier becomes 
stronger with angular momentum.

It is useful to study the density distribution across the shock since 
cooling processes as well as emergent radiation from the disc are strongly 
depends on it \citep{cm06}. For this we compute compression ratio $(R)$ 
which is defined as the ratio of the vertical averaged post-shock 
density to the pre-shock density. As an example, in Fig. 6b, we show the variation of 
compression ratio with the cooling parameter for the same set of flow parameters as in Fig 6a.
For a given set of local energy and the angular momentum of the flow, compression ratio $R$
increases monotonically with the cooling parameter. 
In a convergent accretion flow, shock is pushed inwards when cooling
is enhanced. This causes more compression in the post-shock region. There is a cut-off at a 
critical cooling limit as shock disappears there.

In our further study of shock properties, in Fig. 6c, we draw the variation of the shock
strength ($\Theta$) with the cooling parameter.   Shock strength 
is defined as the ratio of the pre-shock Mach number to the post-shock Mach number and it 
determines the temperature jump at the shock. Same set of initial parameters as in  
Fig. 6a are used here. In the dissipation free limit, the strength of the
shock is weak and it becomes stronger as cooling parameter is increased.

\subsection{General behavior of Shocks}

\begin{figure}
\begin{center}
\includegraphics[height=10cm,width=10cm]{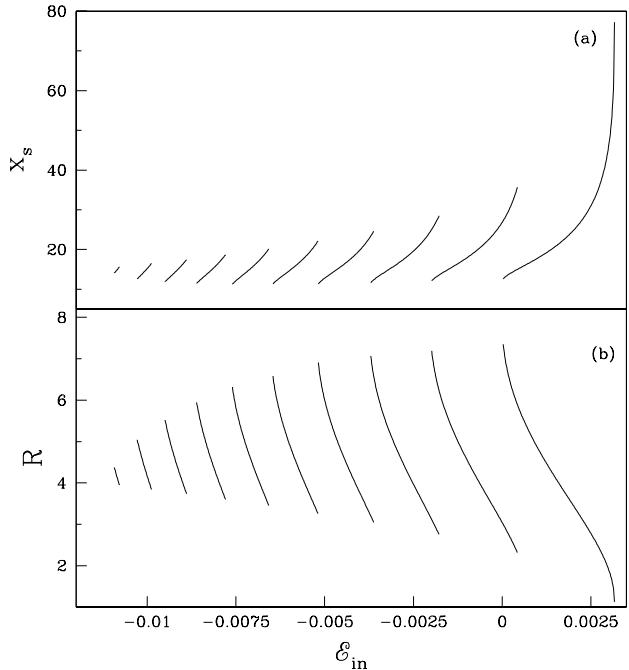}
\caption{\small (a) Variation of shock location with flow energy at the inner
sonic point. The different curves are drawn for different cooling factor.
The rightmost curve is drawn for dissipation free flow. As dissipation
increases $(\Delta \beta = 0.016)$, shock forms at a closer distance. 
(b) Variation of compression ratio as a function of flow energy at the 
inner sonic point and the cooling factor. The flow parameters are same as
(a). Compression ratio decreases
in both higher and lower ends as the cooling is enhanced.
}
\end{center}
\end{figure}

So far, we only study the properties of shock waves when matter is injected from a
fixed outer edge of the disc. In reality, global solutions that contain shock waves 
are not isolated solutions, but are present in a large range of the flow parameters.
We continue our study of shock properties for a wide range of flow parameters. In
Fig. 7a, we draw the shock location $(x_s)$ as a function of specific energy at the
inner sonic point $\mathcal {E}_{in}$.  
The different curves are drawn for a set of cooling parameters starting form
$\beta = 0.0$ (rightmost) to $0.144$ (leftmost) with an interval $\Delta \beta = 0.016$.
Here, the angular momentum at the inner sonic points is chosen as $\lambda_{in}=1.75$. 
Notice that shock forms only for a particular range of $\mathcal {E}_{in}$ and
in the weak cooling limit, shock location is varied in a wide range 
(from 10 - 80 $r_g$ in this particular set of parameters). 
At higher cooling parameter $\beta$, shocks form at closer radii 
and the range of shock locations gradually shrinks. In particular, the upper
limit of the shock locations proceeds more rapidly towards the black hole 
horizon with the increase of cooling parameter.
When $\beta > \beta_{cs}$ is reached, accreting flow fails to satisfy the 
RHC and shock disappears. Shocks can even form at negative energies at the 
inner sonic point (in general, the flow is supposed to have positive energy
at the shock) indicates that due to cooling, flow looses
significant amount of energy while falling into the black hole.

In Fig. 7b, we plot the variation of the compression ratio with $\mathcal {E}_{in}$
for the same set of initial flow parameters used in Fig. 7a. Each individual
curve is obtain for a given cooling parameter ranging from $\beta = 0.0$ (rightmost)
to $\beta = 0.144$ at intervals of $\Delta \beta = 0.016$. 
Strongest shocks ($R \rightarrow 7$) form only in the weak cooling limit.
Figure clearly shows that the upper and lower limit of compression 
ratio monotonically decrease for the gradual increase of the cooling parameter 
and finally compression merges to the intermediate value ($R \sim 4$, 
a measure of strong shock, \citealt{dcnc01} and references there in) upto a critical 
cooling limit beyond which shock does not form.   
Thus, strong shocks are formed for a large range of cooling parameter and outflow 
and jet are obviously expected to be produced in presence of synchrotron cooling.

\section{Parameter Space for shock}

\begin{figure}
\begin{center}
\includegraphics[height=10cm,width=10cm]{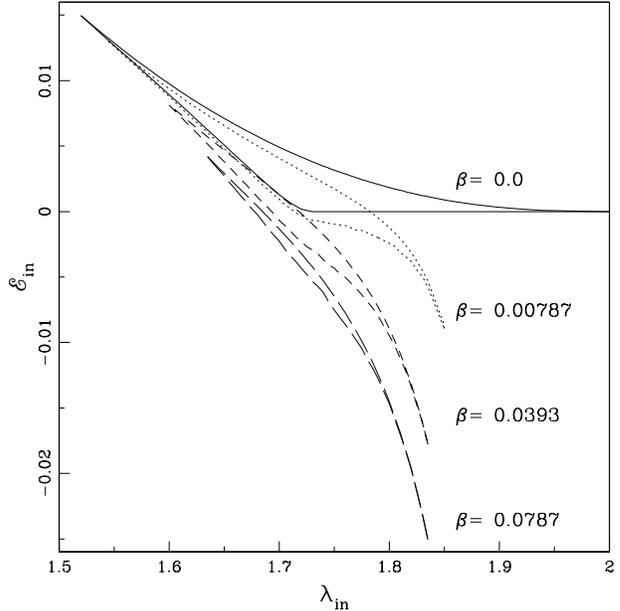}
\caption{\small Modification  of Parameter space for standing shock as a function
of cooling parameter. Parameter space shrinks with 
the increase of cooling factor.
}
\end{center}
\end{figure}

In Fig. 8, we redraw the parameter space as in Fig. 2, but consider the formation
of standing shock only in a cooling dominated flow. The parameter space is
expected to be modified with the increase of cooling parameter. We identify the 
modified regions of the parameter space for stationary shock using various
cooling parameter. We write cooling parameters in the figure. For $\beta = 0.0$, the
region bounded by the solid curve identically merges with that of \citet{c89}.
At higher cooling parameter, the effective region of the parameter space
for standing shock is reduced in both the higher and lower angular momentum
side and it is shifted to negative energy region as well. Beyond a critical
cooling limit this region disappears completely.   

\begin{figure}
\begin{center}
\includegraphics[height=9cm,width=9cm]{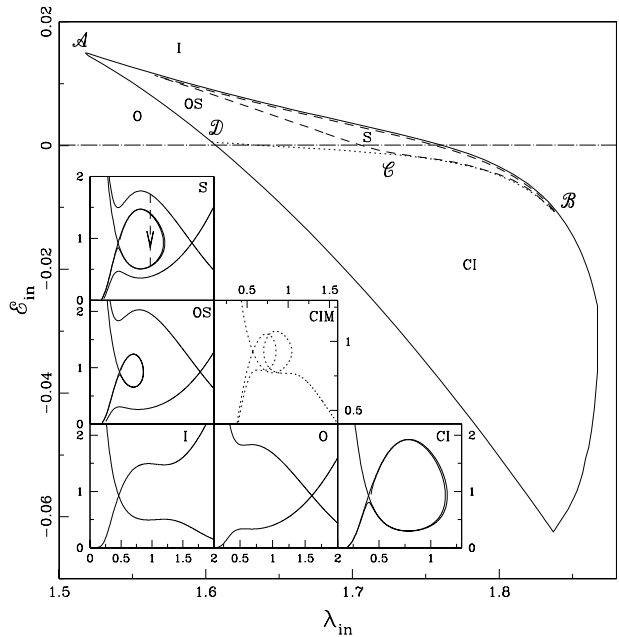}
\caption{\small Division of parameter space in the ${\mathcal E}_{in}, 
\lambda_{in}$ plane according to the topologies. Accretion solutions with
parameters taken from different parts of the parameter space are shown
in the insets. 
}
\end{center}
\end{figure}

We continue our study of parameter space in the energy-angular momentum 
(${\mathcal E}_{in}, \lambda_{in}$) plane according to the accretion flow 
topologies for $\beta = 0.00787$ in the weak viscosity limit. The boundary
denoted by the solid curve in Fig. 9 identifies the region of the 
parameter space for closed topologies passing through the inner sonic 
point (Fig. 1). Further sub-divisions of parameter space are marked by dashed and dotted
curves depending on the behavior of the solution topologies.  Accretion 
topologies with parameters chosen from different region of the parameter 
space are presented in the small boxes at the bottom left of Fig. 9. In each
small boxes, we plot the variation of Mach number as a function of
logarithmic radial distance. Flows with parameters from the region
$\mathcal {ABD}$ possess multiple sonic points.
In particular, when the flow parameters fall in
the region $\mathcal {ABC}$ separated by the dashed boundary, 
RHCs are satisfied. Accretion solution of this kind
is drawn in box labeled S. But, flows with parameters from rest 
of the region of $\mathcal {ABD}$, RHCs are not satisfied, however, the entropy of 
the flow at the inner sonic point continues to be higher compared 
to that at the outer sonic point. In this case, shock starts oscillating causing 
a periodic breathing of post-shock region \citep{rcm97}. The box marked OS 
shows an accretion solution having oscillating shock. For higher cooling, we
obtain a new solution topology as shown in box CIM as seen in \citet{dc04}.
We draw this solution with dotted curve as it is obtained for higher cooling parameter.
This multi-valued solution is expected to be unstable and may cause a non-steady 
accretion. The accretion solution with parameters from the I region of the
parameter space is drawn in the box labeled I. This accretion solution straight away
passes through the inner sonic point before entering into the black hole.
Solution inside the box O represents an accretion flow which passes
through the outer sonic point only. The box marked CI shows an closed accretion solution
passing through the inner sonic point. This type of solution does not extend to the 
outer edge to join smoothly with any Keplerian disc and flow is expected to 
be unstable. When dissipation is significantly increased, the closed topology of CI
opens up (Fig. 1) and then flow can join with the Keplerian disc.

\begin{figure}
\begin{center}
\includegraphics[height=10cm,width=10cm]{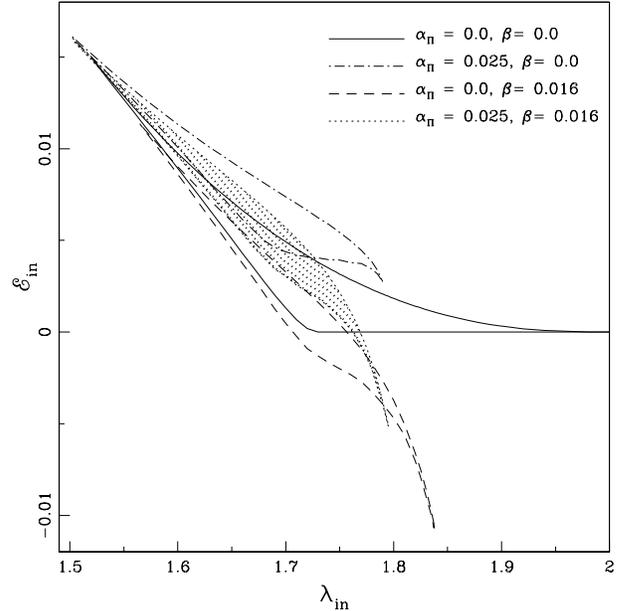}
\caption{\small Region of the parameter space for standing shock for
different dissipation parameters. Dissipation parameters are marked in the figure. 
}
\end{center}
\end{figure}

Fig. 10 shows the comparative study of the parameter space for standing shocks
in different dissipation limits. The dissipation parameters are marked. The solid 
boundary separates the parameter
space for standing shock in the non-dissipative accretion flow \citep{c89}. 
When the viscosity is included, the effective region of the parameter space for standing 
shock reduces and shifts towards the higher energy and the lower angular momentum 
regime \citep{cd04}. The region bounded by dot-dashed curve is obtained for standing
shocks in viscous flow. On the other hand, for a cooling dominated inviscid accretion flow
parameter space for standing shock shrinks in both ends of the angular momentum 
(lower and higher) and moved to the lower energy sides (present paper). This  
parameter space is drawn with dashed curve. When viscosity and cooling are
both present, the parameter space for shock settles down at a intermediate region
(shaded part). This indicates that viscosity and Synchrotron cooling
act oppositely for deciding the shock parameter space. However,
they behave similarly while controlling the dynamics of the shock waves (Fig. 4). 

\section{Critical cooling parameter}

\begin{figure}
\begin{center}
\includegraphics[height=10cm,width=10cm]{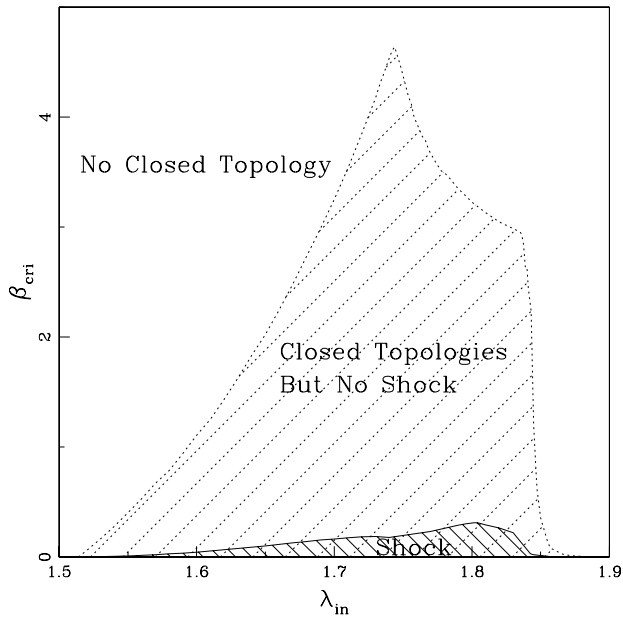}
\caption{\small Variation of critical cooling parameter as a function of angular
momentum at the inner sonic point. The region outside the dotted curve allows 
solutions with open topologies and the region shaded by solid lines separating
the standing shock solutions from the oscillating shock.
}
\end{center}
\end{figure}

In our earlier discussion, we already mentioned that the global behavior of the cooling 
dominated accretion flow changes when the cooling parameter exceeds its critical value.
In particular, there exists two critical cooling parameters. The first one identifies
a boundary which separates the closed topologies from the open topologies passing through
the inner sonic point. The second one is obtained at the region of the closed solutions 
based on the criteria of whether standing shock is formed or not. These critical cooling 
parameters of course strongly depend on the inflow parameters. In Fig. 11, we present
the plot of critical cooling parameters against angular momentum at the inner sonic point.
Different regions are marked. Figure shows that closed topologies as well as shocks are 
formed in the lower and higher angular momentum side for smaller cooling parameters. As 
cooling parameter is increased, shock forms only somewhere at the intermediate angular 
momentum domain which is consistent with our earlier discussion. 

\section{Concluding Remarks}

In this paper, we have presented the properties of the viscous accretion
flows around a stellar-mass black hole in presence of 
synchrotron cooling.
In a realistic accretion flows, the energy loss due 
to radiation is not negligible
and it has significant effect on the dynamical structure of the flow as we 
obtain here.  This will alter the emitted spectrum and luminosity as well.

In the past, efforts were made to study the properties of viscous transonic flow in the
cooling free limit (\citealt{c90a}, \citealt{cd04}) 
and/or cooling is considered in an approximate way \citep{c96b}. No attempts were 
made to study a complete accretion solution which may contain 
shocks in presence of synchrotron cooling. Here, we have obtained a complete set of global 
solutions. We showed that for a large region of the parameter
space a transonic flow can have shock waves when viscosity as well as
synchrotron cooling is very high. We find that the shock 
locations may vary from around ten Schwarzschild radii to several tens of 
Schwarzschild radii depending on the inflow parameters.
We showed that standing shocks where significant 
amount of kinetic energy to thermal energy conversion takes place
can form much closer to the black hole with the enhancement of dissipation. For higher
viscosity, flow transports angular momentum more and more, as the flow moves inward and at the
same time it tends to heat the flow. This causes the centrifugal pressure to be 
much lower at the post-shock region although the post-shock thermal pressure
is higher. When the viscosity parameter is increased, shock forms closer to 
the black hole horizon, indicates that perhaps shocks are centrifugal pressure supported.
On the other hand, Synchrotron photons generated by the stochastic/random magnetic
field can cool down the flow and in addition, the flow angular momentum distribution remains 
insensitive to it (Eq. 3). As the synchrotron cooling is increased, the post-shock 
flow is cooled down more efficiently and the thermal pressure in this region 
reduces significantly. This causes the shock to move inwards to maintain 
pressure balance (thermal plus ram) across it. Thus, viscosity and cooling 
play similar role as far as the dynamic properties of the shock is
concerned. However, cooling can not completely  nullify the effect of viscous 
heating as they depend differently on the flow variables.

We obtain the parameter space for standing shock waves as a function of
cooling parameter and find that the effective region of the parameter space shrinks 
as the cooling parameter is enhanced. We also identify the region of the 
parameter space where standing
shock conditions (RHCs) are not favourable but accreting solutions having two 
saddle type sonic point are present. These solutions generally exhibits 
non-steady shocks \citep{rcm97} and it
could even form when cooling is very high. Moreover, the possibility of 
shock formations decreases for increasing cooling parameter and when
cooling parameter is increased beyond a critical value, shocks disappear 
(Fig. 11).
We also find that viscosity and Synchrotron cooling induce completely opposite effect
in deciding the parameter space for stationary shock waves. 
Indeed, it is clear
that cooling does affect the dynamics of the accretion flow as well as shock 
parameter space quite significantly. Therefore, cooling free assumption \citep{gl04} is
perhaps too simplistic in a general transonic accretion flow.

We have discussed that the black hole accretion models may 
need to include shock waves as they provide a complete explanation of the
observed steady state, as well as time dependent behaviour (\citealt{ct95},
\citealt{rcm97}). 
In particular, as cooling is increased, post-shock energy
goes down and shock location is reduced. In addition, earlier 
studies (\citealt{cm00}, \citealt{rcm97})
show that quasi-periodic oscillations of hard radiations 
from black holes is proportional to the infall time and therefore, QPO frequency
increases with the enhancement of cooling parameter. Observational findings do support
this view that for higher accretion rate, QPO frequency increases. Therefore, 
possibly shocks may be an active ingredient in the accretion flow.  

In the present paper, we have shown that for a given set of flow parameters
there exist two critical cooling parameters which separates the parameter
space into three regions (Fig. 11). In the first region, flow passes through a single X-type
sonic point. In region two, flow has multiple
sonic points but Rankine-Hugoniot relations are not favorable anywhere 
while in the third region Rankine-Hugoniot relations are satisfied and 
standing shocks form.

An important point we need to discuss is that since the basic physics of 
advective flows are similar around a neutron star
(differs from black hole solution only due to inner boundary conditions),
the conclusion may remain roughly similar although the boundary 
of the neutron star, where half of the binding energy could be released, 
may be more luminous than that of a black hole.

In our calculation, we have made some approximations, such as  we consider
stochastic magnetic field instead of large scale field. We ignore the 
effect of radiation pressure. Also, we assume a fixed polytropic index 
rather than computing it self-consistently. We use Paczy{\' n}ski-Wiita 
pseudo-potential \citep{pw80} to describe the space-time geometry around a 
Schwarzschild black hole. However, we believe that our basic conclusion 
will not be affected qualitatively by these approximations. More generalized 
calculations around a rotating black hole is beyond the scope of the 
present paper and will be reported elsewhere.

\section*{Acknowledgments}
The author is thankful to Sandip K. Chakrabarti for his useful suggestions
and discussions. He also thanks I. Chattopadhyay for suggesting the
improvement of the manuscripts.
This work was supported by KOSEF through Astrophysical Research Center
for the Structure and Evolution of the Cosmos(ARCSEC).

\end{document}